\documentclass[aps,prd,twocolumn,groupedaddress,showpacs]{revtex4}
\usepackage{graphicx}
\usepackage{dcolumn}
\usepackage{bm}
\usepackage{amssymb,amsmath,latexsym,amsfonts}

\begin{document}

\title{Planck Scale Induced Speed of Sound in a Trapped Bose--Einstein Condensate}

\author{E. Castellanos\footnote{Dedicated to the loving memory of my Father, El\'ias Castellanos de Luna, RIP 2014.\\
}}
\email{ecastellanos@fis.cinvestav.mx} \affiliation{Departamento de
F\'isica,\\
 Centro de Investigaci\'on y de Estudios Avanzados del
Instituto Polit\'ecnico Nacional, A. P. 14-740, 07000 M\'exico D.
F., M\'exico.}
\author{ J. I. Rivas}
\email{jirs@xanum.uam.mx} \affiliation{Departamento de F\'isica, Universidad Aut\'onoma
Metropolitana-Iztapalapa,\\
A. P.  55-534, 09340 M\'exico D.
F., M\'exico.}

\author{V. Dom\'inguez-Rocha}
\email{vdr@xanum.uam.mx} \affiliation{Departamento de F\'isica, Universidad Aut\'onoma
Metropolitana-Iztapalapa,\\
A. P.  55-534, 09340 M\'exico D.
F., M\'exico.}

\begin{abstract}
In the present work, we analyze the corrections caused by an anomalous dispersion relation, suggested in several quantum gravity models, upon the speed of sound in a weakly interacting Bose--Einstein Condensate, trapped in a potential of the form $V(r)\sim r^{2}$. We show that
the corresponding ground state energy and consequently, the associated speed of sound, present corrections respect to the usual case, which may be used to explore the sensitivity to Planck--scale effects on these relevant properties associated with the condensate. Indeed, we stress that this type of macroscopic bodies may be more sensitive, under certain conditions, to Planck--scale manifestations than its constituents. In addition, we prove that  the inclusion of a trapping potential, together with many--body contributions, improves the sensitivity to Planck--scale signals, compared to the homogeneous system.
\end{abstract}


\maketitle

\section{Introduction}

In some quantum--gravity schemes, the possibility that the
space-time could be quantized, can be characterized, from a
phenomenological point of view, as a modification in the
dispersion relation of microscopic particles
\cite{ami,Giovanni1,Kostelecky,amelino1,Claus,Claus1,5,12,13}
(and references therein). This characteristic feature, emerges
as an adequate tool to test experimentally some quantum--gravity
effects. Nevertheless, the most difficult aspect in searching
experimental hints relevant for the quantum--gravity problem is the
smallness of the involved effects \cite{Kostelecky,amelino1}. If
this kind of deformations are characterized by some Planck scale,
then the quantum gravity effects becomes very small
\cite{Giovanni1,Claus}.

In the non--relativistic limit, it is generally accepted that the
deformed dispersion relation for the energy $\epsilon$ of microscopic particles can be
expressed, in ordinary units, as follows \cite{Claus,Claus1}
\begin{equation}
\epsilon \simeq
mc^2+\frac{p^{2}}{2m}+\frac{1}{2M_{p}}\Bigl(\xi_{1}mcp+\xi_{2}p^{2}+\xi_{3}\frac{p^{3}}{mc}\Bigr),
\label{ddr}
\end{equation}
being $c$ the speed of light, and $M_{p}$ ($\simeq 2.18 \times
10^{-8} Kg$) the Planck mass. The three parameters $\xi_{1}$,
$\xi_{2}$, and $\xi_{3}$, are model dependent
\cite{Giovanni1,Claus}, and should take positive or negative values
close to $1$. There is some evidence within the formalism of Loop
quantum gravity \cite{Claus,Claus1,5,12} that indicates non--zero
values for the three parameters, $\xi_{1},\, \xi_{2},\, \xi_{3}$,
and particularly \cite{5,13} that produces a linear--momentum term
in the non--relativistic limit.

On the other hand, in Refs. \cite{Claus,Claus1} it was suggested the use of
ultra--precise cold--atom--recoil experiments to constrain the form
of the energy-momentum dispersion relation in the non--relativistic
limit. There, the bound obtained for $\xi_{1}$ is at least four
orders of magnitude smaller than the corresponding bound obtained in
Ref. \cite{Castellanos} associated with the condensation
temperature in a Bose--Einstein condensate, trapped in a harmonic
oscillator potential. However, the results obtained in reference
\cite{Castellanos}, suggest that many--body contributions would
allow to improve, in principle, the bounds associated with
$\xi_{1}$. In other words, many--body systems could be used also
to test the sensitivity for some effects arising from Planck scale
regime.

Here, let us add that among the issues related to Bose--Einstein
condensates, we find its possible use as tools in searching
quantum and also classical gravity manifestations, for instance, to detect gravitomagnetic effects, in the context of
Lorentz violation or to provide phenomenological constrains on
Planck--scale physics,
\cite{Camacho,CastellanosCamacho,CastellanosCamacho1,CastellanosClaus,Eli1,Castellanos,CastellanosCamacho0,cam1,r1,r2,echa}.

Indeed, many body systems as tools in searching quantum gravity manifestations, through modifications of the uncertainty principle, has been analyzed in the context of the center of mass motion of macroscopic bodies \cite{I,JD,AM}.
However, in Ref. \cite{AM}, it was argued that this feature can not be used in this context, due to an \emph{incorrect extrapolation criterion} of Planck--scale space--time quantization for fundamental particles, to macroscopic bodies. According to Ref. \cite{AM}, the center of mass motion of a macroscopic body should be affected by Planck--scale quantization, more weakly than its constituents due to a suppression of the form $N^{-s}$, being $N$ the number of particles composing the system, with $s$ some positive power.

On the other hand, the use of the Bogoliubov formalism associated with a
Bose--Einstein condensate confined in a box, open the possibility to explore some alternative scenarios associated with
Planck scale manifestations in the corresponding ground state energy
of the N--body system, by analyzing the associated speed of sound,
as it was reported in Ref. \cite{Eli1}.
There, it was proved that the corrections upon the ground state energy and the corresponding speed of sound, caused by a deformed dispersion relation, scales as a non trivial function of the number of particles (or the corresponding density), together with some parameters associated with the trap. Thus, the argument that a macroscopic body should be affected by Planck--scale quantization more weakly than its constituents, does not seems to be a generic criterion.

Furthermore, let us emphasize that the approach followed in Ref. \cite{Eli1}, and also the followed in the present analysis, suggest alternative scenarios compared to those followed in Refs. \cite{I,JD,AM}. Basically, the main difference lies in the in the fact that the corrections caused by a deformed dispersion relation represents a collective behavior of all the particles forming the condensate, not the properties of a single point, like the center of mass motion. In other words, the corrections caused by the deformation parameters are analyzed over some properties of the condensate, in which, the corrections caused by a deformed dispersion relation on the ground state energy and the corresponding pressure (see below), scales with the number of particles, together with a non--trivial function of the trap parameters. Additionally, the approach followed in the present manuscript (and also the results obtained in
Ref. \cite{Eli1}) suggests that this type of macroscopic systems, \emph{i.e.}, a Bose--Einstein condensate, may be more sensitive, in some cases, to Planck--scale manifestations than its constituents.

Let us remark that the analysis made in Ref. \cite{Eli1}
corresponds to a Bose--Einstein condensate confined in
a box, and it is clear from the experimental point of view, that
there is no condensate in a box. Notice that if non--universal effects are taken into account then corrections of higher order in the ground state energy, and consequently in the corresponding speed of sound, can be obtained \cite{B1,B2,Jo}. These corrections were also compared with lattice Monte Carlo simulations in Ref. \cite{SG}, for a homogeneous condensate. These contributions arise from the fact that non--universal effects, which are sensitive to 3-body physics, are taken into account. However, these systems does not involve more realistic trapping potentials.

Usually, the confinement of the
condensate can be obtained by using harmonic traps, among others
\cite{Dalfovo}. In order to extend our analysis to a more realistic
scenario, the formalism developed in Ref. \cite{Eli1}
must be also extended to a Bose--Einstein condensate trapped in a
harmonic potential that we assume in this report, for simplicity, as $V(r)\sim r^{2}$.

In fact, as we will see later in the manuscript,
the inclusion of a trapping potential improves the sensitivity to Planck--scale effects of the macroscopic system, upon the corresponding speed of sound, in almost four orders of magnitude, compared to the sensitivity obtained in Ref. \cite{Eli1}, for a Bose--Einstein condensate in a box. Thus, we have also an additional tool in this scenario, \emph{i.e.}, a many--body system plus the inclusion of a trapping potential, in which both properties could be used, in principle, to improve the sensitivity of the system  to Planck--scale effects.

In this aim, we define the following  $N$-body
\emph{modified} Hamiltonian which describes our system
\begin{eqnarray}
 \label{MH}
 \hat{H}&=&-\frac{\hbar^{2}}{2m}(1+2m \alpha_{2})\sum_{\delta,\gamma}\langle\delta\vert\nabla^{2}\vert \gamma\rangle\hat{a}^{\dag}_{\delta}\hat{a}_{\gamma}
 \\ \nonumber
 &+&\frac{1}{2} \sum_{\delta,\gamma,\mu,\nu}\langle\delta,\gamma\vert V_{int}(\vec{r})\vert\mu,\nu\rangle
 \hat{a}^{\dag}_{\delta}\hat{a}_{\gamma}\hat{a}^{\dag}_{\mu}\hat{a}_{\nu}
 \\ \nonumber
 &+& \sum_{\delta,\gamma}\langle\delta \vert V_{ext}(\vec{r})\vert\gamma\rangle\hat{a}^{\dag}_{\delta}\hat{a}_{\gamma}
 \\ \nonumber
 &+&\hbar \alpha_{1}\sum_{\delta,\gamma}\langle\delta\vert\sqrt{\vert\nabla\vert^{2}}\, \vert\gamma\rangle\hat{a}^{\dag}_{\delta}\hat{a}_{\gamma}+ \sum_{\delta,\gamma}  mc^ {2}\langle\delta \vert \gamma\rangle\hat{a}^{\dag}_{\delta}\hat{a}_{\gamma}, \,\,\,
\end{eqnarray}
where the operators $\hat{a}$ and $\hat{a}^\dag $, correspond to the
creation and annihilation operators for bosons, satisfying the usual canonical commutation relations
\begin{equation}
[\hat{a}_{\mu},\hat{a}_{\nu}^{\dagger}]=\delta_{\mu \nu},\,\,\,\ [\hat{a}_{\mu},\hat{a}_{\nu}]=[\hat{a}_{\mu}^{\dagger},\hat{a}_{\nu}^{\dagger}]=0.
\end{equation}

Notice also that we
have included the leading order modification in the deformed
dispersion relation (\ref{ddr}), through the \emph{linear} operator
$\vert\sqrt{\vert\nabla\vert^{2}}\vert$, being
$\alpha_{1}=\xi_{1}\frac{mc}{2M_{p}}$, together with the corrections due to the next leading term $\alpha_{2}=\frac{\xi_{2}}{2M_{p}}$. Additionally, $mc^{2}$ is the rest energy.
Clearly, if we set $\alpha_{1}=\alpha_{2}=0$, we recover the usual N--body
Hamiltonian \cite{Ueda}.  

The term, $V_{int}(\vec{r})$ denotes the inter--particle
potential, that will be assumed as  $ V_{int}(\vec{r})\equiv U_{0}=\frac{4 \pi
\hbar^{2}}{m}a$, with $a$ the s--wave scattering length, \emph{i.e.}, at low temperature, only two--body interactions are taken into account. In other words, the system is diluted, and fulfills the condition $n|a|^{3}<<1$, where $n$ is the density of particles \cite{Dalfovo,Pitaevski,Pethick}.
Additionally, $V_{ext}(r)=\frac{m \omega^{2}}{2}r^{2}$ depicts
the trapping potential, where $\omega$ is the corresponding frequency.

The main goal of this work is to analyze, with a simple and compact procedure, the corrections caused by a
deformed dispersion relation on the properties associated with
the ground state energy of a weakly interacting Bose--Einstein condensate. We show that the inclusion of a trapping potential, improves the possible Planck scale signals, compared to those obtained for the homogeneous system. We analyze the
corresponding speed of sound, in order to explore the sensitivity of the system to Planck scale signals. We restrict ourselves to universal effects and to the leading order modification caused by the s--wave scattering length. This is an appropriated approximation, due to the fact that the system under study is enough diluted. Moreover, these problems could be solved, in principle, just tuning the interaction coupling by Feshbach resonances to very small values of the scattering length $a$, in which only the leading term contribution in this parameter is relevant. Therefore, we consider that only two body interactions are relevant. Nevertheless, higher orders caused by non universal effects in the ground state energy for trapped systems could be relevant in this scenario, and deserves further analysis. Finally, we show that the
many--body contributions associated with this system, plus the inclusion of a trapping potential, could be more sensitive to Planck--scale effects than its constituents under typical
laboratory conditions.

\section{Modified Speed of Sound}
Let us calculate the ground state energy and the corresponding speed
of sound associated with our system.
Notice that in the corresponding N--body \emph{modified} Hamiltonian
(\ref{MH}) we have the following terms
\begin{equation}
\label{ef1}
\langle\delta\vert{\nabla}^{2}\vert\gamma\rangle=\int{d}^{3}r{u}_{\delta}^{*}(\vec{r}){\nabla}^{2}{u}_{\gamma}(\vec{r}),
\end{equation}
\begin{eqnarray}
\langle\delta,\gamma\vert
V_{int}(\vec{r})\vert\mu,\nu\rangle =\int\int{d}^{3}{r}_{1}{d}^{3}{r}_{2}{u}_{\delta}^{*}(\vec{r}_{1}){u}_{\gamma}^{*}(\vec{r}_{2})
\\ \nonumber
\times V_{int}(\vec{r}){u}_{\mu}(\vec{r}_{2}){u}_{\nu}(\vec{r}_{1}),
\end{eqnarray}
\begin{equation}
\langle\delta \vert
V_{ext}(\vec{r})\vert\gamma\rangle=\int{d}^{3}r{u}_{\delta}^{*}(\vec{r})V_{ext}(\vec{r})
{u}_{\gamma}(\vec{r}),
\end{equation}
\begin{equation}
\label{ef2}
\langle\delta\vert\sqrt{\vert\nabla\vert^{2}}\vert\gamma\rangle=\int{d}^{3}r{u}_{\delta}^{*}(\vec{r})\sqrt{\vert\nabla\vert^{2}}{u}_{\gamma}(\vec{r}),
\end{equation}
where $\{{u}_{\gamma}(\vec{r})\}$ is a set of single--particle
functions.

In the context of the calculation concerning to the speed of
sound in a Bose--Einstein condensate, usually two choices are made,
namely, (i) The eigenfunctions of a three-dimensional harmonic
oscillator \cite{Zaremba}; (ii) Free particle  wave--function
\cite{Ueda}. These choices are taken without checking if they
correspond to the minimum required by the Gibbs--Bogoliubov--Feynman
formalism \cite{FE}. Nevertheless, the first choice, for the case of
a weakly interacting Bose--Einstein condensate trapped in an isotropic three-dimensional harmonic
oscillator potential, seems to be a good conjecture since it
reflects the symmetry of the trap.

In order to obtain a more appropriated wave--function for our system, let us propose the following single--particle \emph{modified} Hamiltonian, in which we have inserted the contributions due to the deformation parameters $\alpha_{1}$ and $\alpha_{2}$
\begin{equation}
\label{eq:modH}
\hat{H} = \frac{\hat{p}^2}{2m} + V(\hat{r}) + \alpha_1\hat{p} + \alpha_2\hat{p}^2,
\end{equation}
where $m$ is the boson mass and $V(\hat{r})$ is the external potential defined above. The corresponding \emph{modified} wave--function can be easily obtained from the associated equation of motion in the configuration space, with the result
\begin{equation}
\label{psi_n}
\Psi_n(r)=\frac{1}{2^{n/2}n!^{1/2}} \left(\frac{\beta^2\eta^2}{\pi}\right)^{3/4} h_n(q) e^{-\beta^2\eta^2r^2/2},
\end{equation}
where $h_n(q)$ are the Hermite polynomials \cite{Grad}, and $\eta^2 \equiv {(1+2m\alpha_2)^{-1/2}}$. Additionally, ${\beta}^{2}=\frac{m \omega}{\hbar}$.
We observe that the \emph{modified} wave--function Eqn. (\ref{psi_n}), is independent of the deformation parameter $\alpha_{1}$, and setting $\alpha_2 =0$ the usual wave--function is recovered. 

On the other hand, if we further assume that most of the particles are inside the condensate, that is, in the $\vec{p}=0$ state then, this implies that the number of particles in the excited states is negligible. These last assertions can be expressed as follows
\begin{equation}
N_{0}\approx N, \,\,\,\,\,\,\,\, \sum_{\vec{p} \not=0} N_{\vec{p}} <<N,
\end{equation}
being $N$ the total number of particles, $N_{\vec{p}}$ the number of particles in the excited states, and $N_{0}$ the number of particles in the ground state. Keeping terms up to second order in $\hat{a}_{0}$ and $\hat{a}^{\dag}_{0}$, \emph{i.e.}, $\langle \hat{a}_{0}^{\dag} \hat{a}_{0} \rangle= N$, and setting $\alpha_{2}=0$, we are able to obtain the ground
state energy associated with our N--body system
\begin{eqnarray}
\label{E0}
 E_{0}&=&-\frac{{\hbar}^{2}}{2m}\langle0\vert\nabla^{2}\vert0\rangle N + \frac{1}{2}\langle0,0\vert
 V_{int}(\vec{r})\vert0,0\rangle{N}^{2}\\ \nonumber
 &+& \langle0 \vert
 V_{ext}(\vec{r})\vert0\rangle N  +\hbar\alpha_{1}\langle0\vert\sqrt{\vert\nabla\vert^{2}}\vert0\rangle N +
   mc^ {2}\langle0 \vert 0\rangle N.
\end{eqnarray}
In order to calculate the corresponding integrals associated with the N--body ground state energy, let us consider the \emph{modified} eigenfunctions Eqn. (\ref{psi_n}). Thus, we have for instance
\begin{eqnarray}
\langle0\vert\nabla^{2}\vert0\rangle=\int_{0}^{\infty}\int_{\Omega}^{}\Psi_{0}^{*}(r)\nabla^{2}\Psi_{0}(r){r}^{2}\sin\theta
dr d\theta d\phi,\,\,\,\,
\end{eqnarray}
where $\Psi_{0}(r)={\Bigl(\frac{{\beta}^{2}}{\pi}\Bigr)}^{3/4}{e}^{-{\beta}^{2}{r}^{2}/2}$, when $\alpha_{2}=0$, and an equivalent procedure for the other terms in the ground state energy (\ref{E0}).

Using these facts, we are able to obtain the
N--body \emph{modified} ground--state energy associated with our system when $\alpha_{2}=0$
\begin{eqnarray}
 E_{0}=\frac{3}{2}\frac{{\hbar}^{2}}{m {l}^{2}}N+\frac{U_0}{2(2\pi)^{3/2}}\frac{1}{l^3}{N}^{2}
+m{c}^{2}N-{\alpha}_{1}\frac{2 \hbar}{\sqrt{\pi}l}N,
\end{eqnarray}
where we have used equations (\ref{ef1})--(\ref{ef2}), together with
the \emph{modified} eigenfunctions for the ground state associated with a
three-dimensional harmonic oscillator with spherical symmetry.
Notice also that we must define $l^{3}
=V_{char}=(\frac{\hbar}{m\omega})^{3/2}$ as the characteristic
volume associated with our system. In other words, $V_{char}$ can be
interpreted as the available volume occupied by the condensate
\cite{Yan,yuka}.

From the N--body ground state energy $(E_0)$, we are able to
calculate the corresponding speed of sound $v_s^2
=-\frac{{V}^{2}_{char}}{N m}\frac{\partial P_0}{\partial V_{char}}$,
being $P_0 = -\frac{\partial E_0}{\partial V_{char}}$ the ground
state pressure, with the result
\begin{equation}
 v_{s}=\sqrt{\frac{5\hbar^{2}}{3 m^{2}V_{char}^{2/3}}+\frac{U_0N}{(2\pi)^{3/2}mV_{char}} -{\alpha}_{1}\frac{8 \hbar}{9\sqrt{\pi}m{V}_{char}^{1/3}}}.
\end{equation}

Observe that the corrections caused by the deformation parameter $\alpha_{1}$ upon the ground state energy, and the corresponding pressure, scales with the number of particles, together with a non--trivial function of the trap parameters. Unfortunately, the correction on the speed of sound caused by $\alpha_{1}$ is independent of the number of particles, but it depends on the corresponding mass and the frequency as $(m \omega)^{1/4}$, suggesting that massive bosons and/or higher frequencies would allow increase the effects caused by the Planck--scale regime.

The possibility to obtain a measurable correction associated
with the deformation parameter $\alpha_{1}$ ($\delta
v^{\alpha_{1}}_s$) requires that, if $\Delta (v_s)$ is the
experimental error, then $\Delta v_s <\vert\delta
v^{\alpha_{1}}_s\vert$. Thus, this entails
\begin{equation}
\Delta(v_s)\lesssim \sqrt{\vert \xi_{1}\vert \frac{4(\hbar m
c^{2}\omega)^{1/2} }{9\sqrt{\pi}M_{p}}}. \label{Sp101}
\end{equation}

Under these conditions, an experimental uncertainty $\Delta(v_s)$ of
order $10^{-6}\sqrt{\vert \xi_{1}\vert}\,ms^{-1}$, could be tuned, in
principle, below Planck--scale induced speed of sound in
typical conditions, i.e., $\omega\sim 10^{3}$Hz  and $m \sim
10^{-25}$ Kg for $^{87}Rb$ \cite{Dalfovo,aikawa}. In other words, high precision
measurements are required. Here, it is noteworthy to mention that the
accuracy of the speed of sound measurements is not reported in
the literature, at least in the known one by the authors, and consequently,
we can not estimate bounds for the deformation parameter $\xi_{1}$.
Nevertheless, we are capable to estimate the sensitivity of our system to Planck--scale effects.

In order to estimate the sensitivity of our system to Planck--scale effects caused by the
deformation parameter $\alpha_1$, let us appeal to the high precision
experiments in the context of the speed of
sound measurements \cite{Andrews,Andrews2}.

The speed of sound in a condensate is typically of order
$10^{-3}$m$s^{-1}$ \cite{Andrews,Andrews2}. When $\vert
\xi_{1}\vert\lesssim 1$, we obtain a speed of sound of order of a few parts in
$10^{-6}\,ms^{-1}$ under the experimental conditions mentioned
above, that is, three orders of magnitude smaller than the typical
speed of sound reported in references \cite{Andrews,Andrews2},
but four orders of magnitude bigger than the result obtained in
Ref. \cite{Eli1}, which is notable. Indeed, increasing the product $(m
\omega)^{1/4}$, which implies massive bosons and/or higher
frequencies, allows to improve, in principle, the sensitivity to Planck--scale effects.
For instance, when $\xi_{1}\sim 1$, the product $m\omega$ must be of order $10^{-9}$,
in order to obtain a typical speed of sound of order $10^{-3}$m$s^{-1}$, which for frequencies of order
$10^{3}$Hz, involves a boson mass of order $10^{-12}$kg. Conversely, if $m\sim 10^{-25}$kg  then, $\omega \sim 10^{16}$Hz.

Finally, let us briefly focus on the second modification in the deformed dispersion relation Eqn. (\ref{ddr}) with $\alpha_{2}=\xi_{2}/2M_{p}$, \emph {i.e.}, we analyze the corrections on the speed of sound, caused by the deformation parameter $\xi_{2}$, when $\xi_{1}=0$. Following a similar procedure to get Eqn. (\ref{E0}), we obtain that the modified ground-state energy associated with the N-body system is given by
\begin{eqnarray}
\label{eq:E0_alpha12}
E_{0}&=&\frac{3}{2}\frac{{\hbar}^2}{m {l}^{2}}N  + \frac{U_0}{2(2\pi)^{3/2}}\frac{1}{l^3}{N}^{2} +m{c}^{2}N \\\nonumber &+& \alpha_2\left[ \frac{3}{2}\frac{{\hbar}^2}{ {l}^{2}}N - \frac{3U_0}{4(2\pi)^{3/2}}\frac{m}{l^3}{N}^{2}\right].
\end{eqnarray}
Thus, the corresponding speed of sound when $\alpha_1= 0$ reads
\begin{eqnarray}
\label{v_alpha12}
v_{s}^{2} &=& \frac{5\hbar^{2}}{3 m^{2}V_{char}^{2/3}} + \frac{U_0N}{(2\pi)^{3/2}mV_{char}} \\\nonumber &+& \alpha_2\left[\frac{5\hbar^{2}}{3 mV_{char}^{2/3}} -\frac{3U_0 N}{2(2\pi)^{3/2}V_{char}}\right].
\end{eqnarray}

Notice that the corrections caused by the deformation parameter $\alpha_{2}$ on the ground state energy and the speed of sound scales with the number of particles, together with a non--trivial function of the trap parameters and the interaction parameter $U_{0}$.

In this situation, the experimental precision must fulfills the following condition
\begin{equation}
\label{DeltaV_alpha12}
\Delta(v_s)\lesssim \sqrt{\Bigg|\frac{ \xi_{2}}{2M_{p}} \Bigl[\frac{5\hbar \omega}{3} -\frac{3N U_0}{2(2\pi)^{3/2}}\left(\frac{m\omega}{\hbar}\right)^{3/2} \Bigr]\Bigg|}.
\end{equation}
Indeed, the use of
ultra--precise cold--atom--recoil experiments \cite{Claus1} leads to
a bound up to $\vert \xi_{2} \vert \lesssim 10^{9}$.

Here is interesting to notice that when $\vert \xi_{2} \vert \lesssim 1$, with $\omega \sim 10^{3}$$Hz$, $a\sim 10^{-9}$$m$, $m\sim 10^{-25}$$Kg$, and $N \sim 10^{6}$ particles, an experimental precision $\Delta(v_s)$ of order $1.45 \times10^{-10}\, ms^{-1}$ is required. Additionally, when $N\sim 10^{3}$ the required experimental precision is of order $10^{-12}$$ms^{-1}$. These conditions suggest that a large number of particles could be used to improve the experimental precision, in order to obtain a possible measure of the signals caused by the deformation parameter $\alpha_{2}$. 

Let us analyze the sensitivity of our system due to the deformation parameter $\alpha_{2}$. Since $\hbar \omega /M_{p} <<1$ in typical conditions, such a correction
is extremely small, and the second contribution due to $\alpha_{2}$ in Eqn. (\ref{v_alpha12}) could be dominant, indicating that the interaction among the constituents of the system could be representative.
The above argument indicates that when $\xi_{2}>0$, for $N >10^{3}$ particles, this could leads to a not well defined speed of sound correction. 
Notice that for $N >10^{3}$ particles, $\omega \sim 10^{3}$$Hz$, $a\sim 10^{-9}$$m$, and $m\sim 10^{-25}$$Kg$, the deformation parameter $\xi_{2}$ must be negative, and a well defined correction upon the speed of sound of order $\sim 10^{-10}\, ms^{-1}$ is obtained, when $\xi_{2}\sim -1$. In other words, a large number of particles could be used to improve the sensitivity to Planck scale signals, together with the restricted condition $\xi_{2}\sim-1$.

Conversely, when $\xi_{2} \sim 1$ finite size systems of order $N\sim 10^{3}$ particles, together with frequencies of order $\omega \sim10^{3}$$Hz$, $a\sim10^{-9}$$m$, and $m\sim10^{-25}$$Kg$ are required, in order to obtain a well defined correction upon the speed of sound. For $N \sim 10^{3}$ the corresponding correction upon the speed of sound is of order of a few parts in $10^{-12}$, with $\xi_{2}\sim 1$.

Notice that the corrections caused by the deformation parameter $\alpha_{2}$, could be re--absorbed in the usual term by defining the effective mass $m_{\xi_{2}}\equiv M_{p}m/ (M_{p}+ \xi_{2}m)$, see the Hamiltonian (\ref{MH}). Indeed, a similar analysis can be made in order to obtain the ground state energy and the corresponding speed of sound in this situation. However, we stress here that both approaches are equivalent, \emph{i.e.,} treating the contributions of the deformation parameter $\xi_{2}$ in an independent manner, or re--defining an effective mass. Thus, both approaches lead to the same predictions upon the ground state energy, and consequently on the speed of sound contributions, at least to first order in $\xi_{2}$.

\section{Discussion}

We have analyzed the corrections on the speed of sound in a
weakly interacting Bose--Einstein condensate trapped in an isotropic
harmonic oscillator type potential, caused by a deformed dispersion
relation. 

We have proved that the corresponding speed of sound presents a correction, which depending on the sign of the deformation parameters, could be positive or negative. We have obtained, under typical conditions, a correction in the corresponding speed of sound of order $10^{-6}$m$s^{-1}$, caused by the deformation parameter $\alpha_{1}$, when $\xi_{1}\sim1$, and a correction of order $10^{-10}$m$s^{-1}$ for the second deformation parameter $\alpha_{2}$, when $\xi_{2} \sim -1$ for $N >10^{3}$. Conversely, when $\xi_{2}\sim 1$ for $N \sim 10^{3}$, we obtain a correction of order $10^{-12}$$ms^{-1}$ in the speed of sound. However, as expected, the correction caused by $\alpha_{1}$ dominates over the correction caused by $\alpha_{2}$.

At this point is important to mention that the inclusion of a trapping
potential increases the sensitivity to Planck--scale effects associated with $\xi_{1}$, in almost four orders of magnitude, respect to those obtained in
Ref. \cite{Eli1}. This fact suggests that a generic potential
of the form $V(r)\sim r^{s}$, with $s$ a positive real number, could be used to increase even more,
the sensitivity of our system to Planck--scale manifestations,
in the context of speed of sound measurements.

In addition, let us point it out the following: it is clear that the
calculations done in the present work, does not include the possible
corrections on the speed of sound caused by the particles in
the excited states (finite temperature corrections), or even long range interactions (\emph{e.g.},
anisotropic dipole--dipole and/or spin--spin interactions). The
particles in the exited states and some long range interactions,
could contribute to the speed of sound measurements
\cite{abelsound,abelsound1}, and in consequence, it could be
interesting to explore if such corrections could be used to improve
the sensitivity associated with the deformation parameters obtained in
the context of the present report.

Concerning to the contributions caused by the deformation parameter proportional to $\xi_3 p^{3}$, we must emphasize that the wave function calculated in the present report do not reflect the symmetry of the system when $\xi_{3} \neq 0$, due to higher order derivatives in the equation of motion. This fact suggests that the wave function expressed in the Eqn. (\ref{psi_n}), seems to be not a good approximation in order to calculate the ground state energy, and consequently the corresponding speed of sound. Additionally, we expect that the contributions due to $\xi_3 p^{3}$ will be smaller several orders of magnitude compared to the corrections caused by $\alpha_1$ and $\alpha_2$. However, it will be interesting to analyze the symmetry of the system taken into account the contributions of $\xi_3 p^{3}$ in order to explore the ground state energy, and the speed of sound. Clearly, this is a non--trivial topic which deserves further investigation and that will be presented elsewhere.

Finally, we emphasize that the approach followed in the present manuscript suggests that this type of macroscopic system, could be more sensitive,
in some cases, to Planck--scale effects than its constituents. Nevertheless, this last assertion do not implies that the current technology is capable
to detect such manifestations. However, it is remarkable that many--body contributions in a Bose--Einstein condensate, open the possibility of planning
specific scenarios that could be used, in principle, to test possible effects caused by the quantum structure of space--time.

\begin{acknowledgments}
This work was partially supported by CONACyT M\'exico under
grants CB-2009-01, no. 132400, CB-2011, no. 166212,  and
I0101/131/07 C-234/07 of the Instituto Avanzado de Cosmolog\'ia
(IAC) collaboration (http://www.iac.edu.mx/). E. C. acknowledges
CONACyT for the postdoctoral grant received.  J. I. Rivas
acknowledges CONACyT grant No. 18176. VDR thanks financial support from
CONACyT grant No. 222276, and
to Flores-Huerta A G for her encouragement.
\end{acknowledgments}

\end{document}